\documentclass[aps,preprint,showpacs,superscriptaddress,groupedaddress]{revtex4}  % for double-spaced preprint
\usepackage{graphicx, subfigure}  % needed for figures and subfigure
\usepackage{dcolumn}   % needed for some tables
\usepackage{bm}        % for math
\usepackage{amssymb}   % for math
\usepackage{amsmath}   % for align equations
\everymath{\displaystyle}

\begin{document}

% The following information is for internal review, please remove them for submission
\widetext
%\leftline{Version xx as of \today}
%\leftline{Primary authors: CHI WUN CHOI}
%\leftline{To be submitted to ???}
%\leftline{Comment to {\tt ???@???.???} by xxx, yyy}
%\centerline{\em D\O\ INTERNAL DOCUMENT -- NOT FOR PUBLIC DISTRIBUTION}

% the following line is for submission, including submission to the arXiv!!
%\hspace{5.2in} \mbox{Fermilab-Pub-04/xxx-E}

\title{Adaptive cyclically dominating game on co-evolving networks:
Numerical and analytic results}
\author{Chi Wun Choi} \affiliation{Department of Physics, The Chinese University of Hong Kong, Hong Kong SAR, China}
\author{Chen Xu} \affiliation{College of Physics, Optoelectronics and Energy,
Soochow University, Suzhou 215006, China}
\author{Pak Ming Hui} \affiliation{Department of Physics, The Chinese University of Hong Kong, Hong Kong SAR, China}
\date{\today}

\begin{abstract}
A co-evolving and adaptive Rock (R)-Paper (P)-Scissors (S) game
(ARPS) in which an agent uses one of three cyclically dominating
strategies is proposed and studied numerically and analytically.
An agent takes adaptive actions to achieve a neighborhood to his
advantage by rewiring a dissatisfying link with a probability $p$
or switching strategy with a probability $1-p$. Numerical results
revealed two phases in the steady state. An active phase for
$p<p_{\text{cri}}$ has one connected network of agents using
different strategies who are continually interacting and taking
adaptive actions.  A frozen phase for $p>p_{\text{cri}}$ has three
separate clusters of agents using only R, P, and S, respectively
with terminated adaptive actions. A mean-field theory of link
densities in co-evolving network is formulated in a general way
that can be readily modified to other co-evolving network problems
of multiple strategies. The analytic results agree with simulation
results on ARPS well.  We point out the different probabilities of
winning, losing, and drawing a game among the agents as the origin
of the small discrepancy between analytic and simulation results.
%A closer examination of the small discrepancy between analytic and
%simulation results reveals the different probabilities of winning,
%losing, and drawing a game among the agents.
As a result of the adaptive actions, agents of higher degrees are
often those being taken advantage of.  Agents with a smaller
(larger) degree than the mean degree have a higher (smaller)
probability of winning than losing. The results are useful in
future attempts on formulating more accurate theories.
\end{abstract}

\maketitle

%=======================================================
\section{Introduction}
\label{Introduction}

Agent-based modelling is an important tool for studying the
autonomous actions of individual entities and their
interactions~\cite{review_szabo,review_statphy} in complex
systems.  The interactions often reflect how agents compete,
especially in the context of competing games. Examples of some
extensively studied games are the prisoner's dilemma (PD),
snowdrift game (SG), and stag hunt game
(SH)~\cite{review_szabo,book_evolutionary,book_sh,book_axelrod}.
These are two-strategy games with agents having a choice of two
possible options. In the present work, we focus on the
Rock-Paper-Scissors (RPS)
game~\cite{review_szabo,book_evolutionary,review_rps},
characterized by three strategies that dominate each other {\em
cyclically}.  Depending on the context, the strategy of an agent
can be regarded as his state, character, opinion or species.  The
strategies are related cyclically through: Rock (R) crushes
Scissors (S), Scissors (S) cuts Paper (P), and Paper (P)
covers Rock (R)~\cite{review_szabo,review_rps,review_zhou}.
Despite its simplicity, many phenomena in nature can be described
within the framework of RPS game.  A well-known example is related
to the mating strategy of the common side-blotched lizards
(\textit{Uta stansburiana}), a species of lizards found in the
western coast of North America \cite{rps_lizard}.  Other examples
include phenomena in marine ecological
communities~\cite{rps_marine}, coexistence of different kinds of
microbes~\cite{rps_bacteria, rps_bacteria2,rps_bacteria3}, and in
chemical and biological systems~\cite{review_szabo,
review_rps,review_zhou, rps_grass}. There are phenomena in
economic and social systems, e.g., human decision-making processes
and epidemic diseases, that also involve cyclical dominance and
they can be studied within the RPS
framework~\cite{review_szabo,review_zhou,rps_exp}.

An interesting question is how network
structures~\cite{review_szabo,review_rps} affect the RPS game. The
focus so far has been on static networks, i.e., the links
connecting two competing agents are fixed.  For RPS agents
interact in a square lattice~\cite{tainaka94} with the
loser updating the strategy to be that of the winner, spatial
self-organized patterns emerged.  Szab\'o {\em et al.} studied the
small-world effect on RPS game by replacing a fraction $r$ of the
links in a square lattice by links that connect two randomly
selected agents~\cite{rps_szabo1}.  It was found that two
qualitatively different phases result, depending on the value of
$r$.  Szolnoki and Szab\'o studied the RPS game in Kagome,
honeycomb, triangular, cubic, and ladder-shape
lattices~\cite{rps_szabo2}. They found that while the spatial
dimension of the lattices affects the transitions between
different phases strongly, the clustering coefficient does not.

Going beyond static networks, co-evolving networks have attracted
much attention in recent years~\cite{review_szabo, review_statphy,
review_rps, review_tgross, review_mini}.  In co-evolving networks,
an agent may switch his strategy or alter his competing neighbors
so as to attain an environment that is to his advantage. Such
adaptive actions couple the dynamics of strategy selections and
network evolution.  Co-evolving networks invoking PD, SG, SH games
have been studied~\cite{review_szabo, review_statphy,
review_tgross,review_mini,dasg1,Xu1}.  In particular, the present
work is motivated by the two-option adaptive co-evolving voter
model~\cite{vm} and the dissatisfied adaptive snowdrift
game~\cite{dasg1,dasg2}.  In the co-evolving voter
model~\cite{vm}, there are two opposite opinions competing for
dominance in an initially random regular network and agents prefer
to be surrounded by like-opinion neighbors.  When an agent
interacts with a randomly chosen neighbor of the opposite opinion,
he has a probability $p$ to cut the link to the neighbor and
rewire it to a randomly chosen agent of the same opinion. With a
probability $1-p$, the agent is convinced by the neighbor and
switches to the opposite opinion.  Both actions are rational in
that the agents tend to pursue local consensus. Despite its
simplicity, the phenomena are rich. For values of $p$ below
(above) a critical value, the system evolves into an active (a
frozen) phase in which the network evolution and strategy
selection continue (cease). Similar adaptive actions (i.e.
switching strategies and rewiring the links to dissatisfying
neighbors) were included in the model of dissatisfied adaptive
snowdrift game (DASG)~\cite{dasg1, dasg2}. In DASG, adaptive
actions are taken when agents become dissatisfied with
non-cooperative neighbors.  The resulting network is either in a
disconnected, dynamically frozen, and character-segregated phase
or a connected, dynamical, and character-mixed phase, depending on
a payoff parameter.  Analytic approaches to co-evolving networks
require careful treatment of spatial correlations~\cite{ji}. Other
examples in which similar adaptive actions are invoked include a
reversed opinion-formation model~\cite{reverse} and an inverse
voter model~\cite{inverse,CWChoi}.  Networking effects, including
co-evolving networks, also pose challenging questions to analytic
approaches. Typically approaches such as mean field approximation
and pair approximation~\cite{review_szabo, review_rps} often only
give results in qualitative agreement with
simulations~\cite{review_szabo, vm, dasg1, dasg2, inverse,
ji,CWChoi}. The reason is that the adaptive actions are sensitive
to the local competing environment and thus spatial correlations
are important. We have made various attempts in understanding the
key factors in formulating theories that better capture spatial
correlations~\cite{dasg2,ji,CWChoi,WZhang1,WZhang2,Elvis}.  An
improved mean field theory was shown to give good results for
DASG~\cite{ji} and the inverse voter model~\cite{CWChoi}.

Here, we generalize the study of adaptive co-evolving models to
cyclic multiple-strategy case.  In particular, an adaptive and
co-evolving RPS model, abbreviated as APRS, is proposed and
studied in detail. Our model is different from the adaptive RPS
model studied by Demirel~{\em et al.}~\cite{arps}. The agents in
adaptive RPS model prefers to have neighbors of the same option by
an adaptive mechanism in which an agent who lost a RPS game adopts
the strategy of the winner or to seek a new neighbor of the same
opinion.  The authors focused on the time evolution of the
fractions of agents using the different strategies.  In our model,
the agents take adaptive actions to enhance their chance of
winning.  We focus on the different phases exhibited in the steady
state and the formulation of analytic approaches.
In Sec.~2, we define our model and identify the key features as
revealed by simulations.  The model is parameterized by a
probability $p$ of rewiring an unfavorable link.  The system
evolves to two different phases for different ranges of $p$.  In
Sec.~3, a theory based on the densities of different kinds of
links connecting agents of different strategies is constructed.
Results are found to be in good agreement with simulations, with
small yet noticeable discrepancies.  In Sec.~4, we point out that
the small discrepancies are important hints for studying the
validity of the assumptions in a theory.  We analyze the
dependence of the probabilities of winning and losing of different
types of agents. These probabilities are found to depend on the
role of an agent in an adaptive process and his degree.  These
features are usually not included in analytic approaches. Although
the context of ARPS is studied, the discussions on the formalism
of mean field theory and its validity are intentionally carried
out in a general form.  As such, the analysis here can be readily
applied to other co-evolving network models with two or more
options or strategies.  Results are summarized in Sec.~5.

%=======================================================
\section{Adaptive Rock-Paper-Scissors model and Key Features}
\label{sec2}

\begin{figure}
\includegraphics[width=0.5\columnwidth]{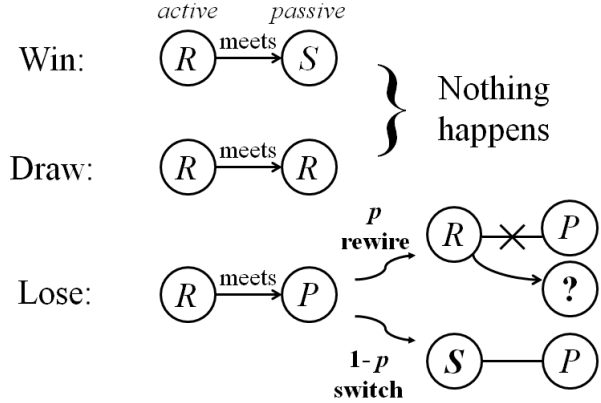}
\caption{Schematic illustrations of the adaptive actions of an
active agent.} \label{fig01}
\end{figure}

Consider a system of $N$ agents.  For concreteness, the agents are
initially connected via a random regular graph of uniform degree
$\mu$ and each of them is assigned one of the three strategies
(R, P, or S) with equal probabilities. In a time step, an
agent, referred to as the {\em active agent}, is selected
randomly. If there is no connected neighbor, i.e. of degree zero,
there will be no action and the time step ends. Otherwise, the
active agent selects a connected neighbor, referred to as the {\em
passive agent}, at random. They interact via a RPS game. If the
active agent wins or there is a draw, he is satisfied and no
adaptive actions take place. If the active agent loses, he is
dissatisfied and he will take one of the following adaptive
actions: (i) with a probability $p$ to cut the link to the passive
agent and rewire it to another agent (called the rewiring target)
randomly chosen from all the agents in the system who are not a
neighbor, or (ii) with a probability $(1- p)$ to switch his
strategy to the one that can defeat the passive agent.
Figure~\ref{fig01} illustrates the possible events and adaptive
actions in a time step with examples. As one active agent is
picked at a time step, the interactions are asynchronous. The
probability $p$ is the only parameter in APRS.  The adaptive
actions are rational in that an agent always aims to prevent
losing to the same opponent by altering the local competing
environment. They drive the strategies employed by the agents and
the network connections to co-evolve. The process continues until
the network achieves a macroscopically steady state.

The long-time behavior of the system can be characterized by a few
macroscopic quantities.  They include the fractions $f_{R}$,
$f_{P}$, and $f_{S}$ of agents using the strategies-R, P, and
S, respectively, the fractions of undirected {\em inert} links
$l_{RR}$, $l_{PP}$ and $l_{SS}$ connecting agents using the same
strategy that would lead to a draw, and the fractions of
undirected active links $l_{RP}$, $l_{PS}$ and $l_{SR}$ connecting
agents using different strategies that would lead to a win-lose
situation.  It should be pointed out that ARPS can be implemented
with different initial strategy assignments and initial network
connections.  Here, we take advantage of the simplicity provided
by random initial strategy assignments and the symmetry among the
three strategies so that we could focus on the discussion of the
$p$-dependence of two link densities, one for inert and the other
for active links.

%=======================================================

\begin{figure}
\subfigure{%Result of $f_R$
\includegraphics[width=0.45\textwidth]{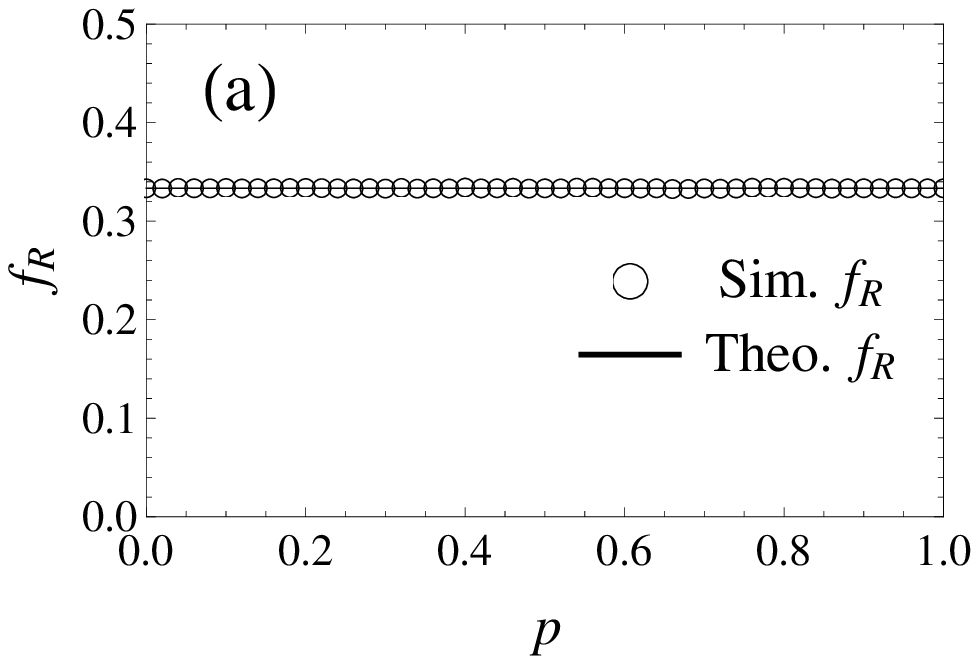}
}%
\subfigure{%Results of $l_{RR}$ and $l_{RP}$
\includegraphics[width=0.45\textwidth]{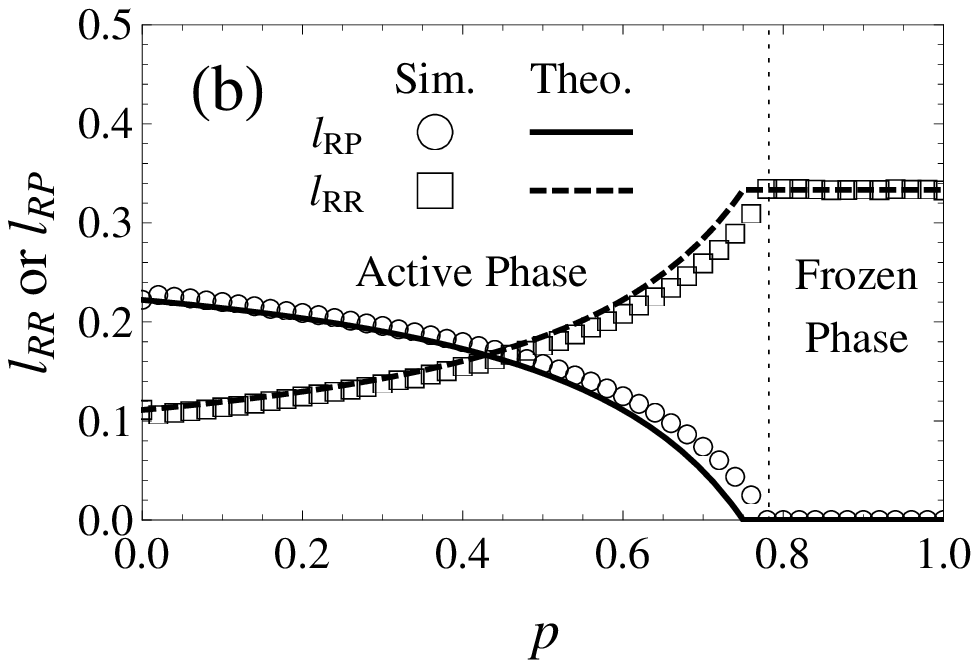}
}%
\caption{%
Simulation (symbols) of (a) $f_R$ and (b) $l_{RR}$ and $l_{RP}$ in
the steady state as a function of the rewiring probability $p$.
The data are obtained by averaging results over 300 independent
runs. The system has $N=10000$ agents and a mean degree $\mu=2$.
The two phases are indicated in (b).  Results of mean field theory
as given by Eqs.(\ref{eq10}) and (\ref{eq11}) are included (lines)
for comparison.
}%
\label{fig02}
\end{figure}

Detailed numerical simulations were carried out for ARPS. Here we
focus on an initial network of uniform degree $\mu=2$ and
$N=10,000$ agents.  The results illustrated that
$f_R=f_P=f_S=1/3$, $l_{RR}=l_{PP}=l_{SS}$ and
$l_{RP}=l_{PS}=l_{SR}$.  These are expected as no strategy plays a
special role in RPS and the adaptive actions in ARPS.  The random
initial conditions make sure that all strategies are evenly
present.  This allows us to focus the discussion on how
$l_{RR}(p)$ and $l_{RP}(p)$ behave at long time. Fig.~\ref{fig02}
shows the simulation results (symbols). Fig.~\ref{fig02}(a)
confirms $f_R=f_P=f_S=1/3$ for all values of $p$, as expected from
symmetry consideration. Fig.~\ref{fig02}(b) shows the behavior of
$l_{RR}(p)$ (squares) and $l_{RP}(p)$ (circles).  These quantities
reveal the two different phases classified by $p$. For
$0<p<p_\text{cri}$, $l_{RR}$ increases monotonically with $p$ and
approaches $l_{RR}=1/3$ at $p=p_\text{cri}$ continuously while
$l_{RP}$ drops monotonically with $p$ and vanishes continuously at
$p=p_\text{cri}$. In the range $p_\text{cri}<p<1$, $l_{RR}=1/3$
and $l_{RP}=0$.  We found that $p_\text{cri} \approx 0.78$ for
$\mu=2$.  We also studied initial networks of different values of
$\mu$.  The results show the same qualitative behavior, but
$p_\text{cri}$ increases with $\mu$. For example, $p_\text{cri}
\approx 0.89$ for $\mu=4$. Here, we focus on analyzing the results
for $\mu=2$.

For $p<p_\text{cri}$, the system has active links and it is in the
{\em active phase}. These active links promote agents'
interactions and adaptive actions.  This is a dynamic phase as
strategy switching and network rewiring persist. For
$p>p_\text{cri}$, the system has only inert links and it is in an
inactive and {\em frozen phase}.  There is no more adaptive
action.  The two phases also differ drastically in network
structure.  In the active phase, the system has a main cluster
consisting of agents using the three strategies with both active
and inert links. In the frozen phase, the system breaks into three
segregated pure-strategy clusters of equal size, with each cluster
having agents using only R, P, or S.  Another noticeable
feature is the discontinuous jump from $l_{RP}=2/9$ at $p=0$ to a
larger value when $p$ becomes finite.  There is a similar
discontinuous jump from $l_{RR} = 1/9$ at $p=0$ to a smaller
value.  These discontinuities will be discussed in Sec.~4.

%=======================================================
\section{Mean-field approach}
\label{sec3}

Inspired by previous analytic approaches~\cite{vm, dasg1, dasg2,
inverse, ji, CWChoi} for co-evolving agent-based models, we
formulate a theory by tracing the expected changes in the
macroscopic quantities in a time step. In principle, the system
has many macroscopic variables.  At the single-agent level, we
have the fractions $f_R$, $f_P$, and $f_S$. At the two-agent or
link level, there are the link densities $l_{RR}$, $l_{PP}$,
$l_{SS}$, $l_{RP}$, $l_{PS}$ and $l_{SR}$.  As discussed, symmetry
implies ${l_{RR}} = {l_{PP}} = {l_{SS}} $ and $ {l_{RP}} =
{l_{PS}} = {l_{SR}}$.  Therefore, we could take $l_{RR}$ and
$l_{RP}$ as variables. Together with ${f_R} = {f_P} = {f_S} =
1/3$, the two variables obey $l_{RR} + l_{RP} = 1/3$.  This sum
rule is also demonstrated by the simulation results in
Fig.~\ref{fig02}(b).  As a result, a single variable suffices for
a theory up to the level of links.  We choose $l_{RP}$ as the
variable, although other choices can also be made.

We formulate a theory in a way that can be readily generalized to
other co-evolving network problems.  To proceed, we aim at writing
down an equation for the change $\Delta l_{RP}$ in $l_{RP}$ in a
time step.  Based on the adaptive actions, $\Delta l_{RP}$ is
determined by: (i) the strategy of the active agent, (ii) his
local configuration including the degree $\kappa$ and the numbers
of neighbors using the different strategies, (iii) the probability
of losing the RPS game, (iv) the adaptive action taken after
losing, (v) the change in the {\em number} of links $\Delta
L_{RP}$ connecting an agent using strategy-R and an agent using
strategy-P (called RP-links) due to the adaptive action.
Table~\ref{tab01} gives the possible values of $\Delta L_{RP}$ due
to the adaptive actions.  Schematically, the expected change in
the link density $\Delta l_{RP}$ can be expressed in terms of the
probabilities of all possible local configurations, strategies and
adaptive actions, and the corresponding local changes in the
number of RP-links as follows:
\begin{eqnarray}
\label{eq01} \Delta l_{RP} = \sum\limits_{X = R,P,S} f_X
\sum\limits_{\kappa} P_X(\kappa)
\sum\limits_{\lambda_{XY},\lambda_{XZ}} Q_{X,\kappa}
(\lambda_{XY},\lambda_{XZ}) \frac{\lambda_{XY}}{\kappa} \left[
\frac{p \Delta L_{RP}^{\text{rewire}}}{L_{\text{total}}} +
\frac{(1-p) \Delta L_{RP}^{\text{switch}}}{L_{\text{total}}}
\right].
\end{eqnarray}
Here, $P_X(\kappa)$ is the probability of an agent using
strategy-X and having degree $\kappa$, Y (Z) is the strategy
which wins over (loses to) X, $Q_{X,\kappa}
(\lambda_{XY},\lambda_{XZ})$ is the probability of an agent using
strategy-X and having degree $\kappa$ to have $\lambda_{XY}$
XY-links and $\lambda_{XZ}$ XZ-links, $\Delta
L_{RP}^{\text{rewire}}$ ($\Delta L_{RP}^{\text{switch}}$) is the
local change in RP-links due to rewiring (switching) with its
possible values listed in Table~\ref{tab01}, and $
L_{\text{total}} = \mu N/2$ is the total number of links in the
network.

\begin{table}[h]
\begin{tabular}{cc} \hline
Adaptive action & Change in RP-links ($\Delta L_{RP}$)\\ \hline
switch strategy from R to S & $-\lambda _{RP}$ \\
switch strategy from S to P & $\lambda _{SR}$ \\
switch strategy from P to R & $\kappa - 2{\lambda _{RP}} - {\lambda _{PS}}$ \\
agent of R cuts P then rewires to R & $-1$ \\
agent of R cuts P then rewires to S & $-1$ \\
agent of P cuts S then rewires to R & $+1$ \\
other actions & $0$ \\ \hline
\end{tabular}
\caption{Changes in the number of RP-links for different
adaptive actions.} \label{tab01}
\end{table}

Eq.~(\ref{eq01}) is general but hard to solve. Formally, the
quantities in the right-hand side changes with time as the
adaptive actions proceed, and dynamical equations tracing their
variations should also be established.  Fortunately,
simplifications are possible when we focus only on the long time
behavior when various quantities become stable in time and close
the equation by proper approximations. Eq.~(\ref{eq01}) can be
written into three terms, each corresponding to the active agent
using X = R, P, S, respectively, i.e.,
\begin{equation}
\label{eq02}
\Delta {l_{RP}} = \frac{2}{{\mu N}}\left( {{f_R}\Delta l_{RP}^R +
{f_P}\Delta l_{RP}^P + {f_S}\Delta l_{RP}^S} \right). \\
\end{equation}
For given strategy-X and value of $\kappa$, there is an expected
value
\begin{equation}
\label{eq03} \sum\limits_{\lambda_{XY},\lambda_{XZ}} Q_{X,\kappa}
(\lambda_{XY},\lambda_{XZ}) (\cdots) \equiv \langle \cdots
\rangle_{\lambda|X,\kappa}
\end{equation}
for the agents using strategy-X and having exactly degree
$\kappa$ to be carried out.  The notation $\langle \cdots
\rangle_{\lambda|X,\kappa}$ stresses two points: (i) the average
is taken over possible $\lambda$'s and (ii) the result is a
function of X and $\kappa$.  Similarly, we further define an
expected value over possible values of the degrees for agents
using strategy-X as:
\begin{equation}
\label{eq04} \sum\limits_\kappa  {{P_X}(\kappa ) ( \cdots )}
\equiv {\langle  \cdots \rangle _{\kappa|X}} \;,
\end{equation}
with the result depending on the strategy-X. The quantities
$\Delta l_{RP}^R$, $\Delta l_{RP}^P$, and $\Delta l_{RP}^S$ in
Eq.~(\ref{eq02}) can be expressed in terms of these expected
values. Explicitly, they can be expressed by using
Table~\ref{tab01} as
\begin{eqnarray}
\label{eq05} \Delta l_{RP}^R & = & - p\left( {{f_R} +
{f_S}}\right){\left\langle {\frac{{{{\left\langle {{\lambda
_{RP}}} \right\rangle }_{\lambda |R,\kappa}}}}{\kappa }}
\right\rangle _{\kappa |R}} - (1 - p){\left\langle
{\frac{{{{\left\langle {{\lambda _{RP}}^2} \right\rangle
}_{\lambda |R,\kappa}}}}{\kappa }} \right\rangle _{\kappa |R}},
\nonumber \\
\Delta l_{RP}^P & = & p{f_R}{\left\langle {\frac{{{{\left\langle
{{\lambda _{PS}}} \right\rangle }_{\lambda |P,\kappa}}}}{\kappa }}
\right\rangle _{\kappa |P}} \nonumber \\
& & + (1 - p) \left( \left\langle \left\langle \lambda
_{PS}\right\rangle_{\lambda |P,\kappa} \right\rangle_{\kappa|P} -
2 \left\langle \frac{{{{\left\langle {{\lambda _{PS}}{\lambda
_{RP}}}\right\rangle}_{\lambda |P,\kappa}}}}{\kappa}
\right\rangle_{\kappa |P} - \left\langle {\frac{{{{\left\langle
{{\lambda _{PS}}^2} \right\rangle}_{\lambda |P,\kappa}}}}{\kappa
}} \right\rangle_{\kappa |P} \right) \,,
\nonumber \\
\Delta l_{RP}^S & = & (1 - p){\left\langle {\frac{{{{\left\langle
{{\lambda _{SR}}^2} \right\rangle }_{\lambda |S,\kappa}}}}{\kappa
}} \right\rangle _{\kappa |S}} \,,
\end{eqnarray}
where the terms proportional to $p$ are due to rewiring and those
proportional to $(1-p)$ are due to strategy switching.

To proceed, we make approximations to the expected values so as to
close the equations.  Firstly, the equations can be simplified by
the symmetry of the three strategies.  As a result, it is
sufficient to consider the expected values in regard to only one
of the strategies. Without loss of generality, we retain averages
over agents using strategy-R.  The other expected values for
strategies-P and S are given by: $\left\langle \left\langle
\lambda _{PS}\right\rangle_{\lambda |P,\kappa}/\kappa
\right\rangle_{\kappa|P} = \left\langle \left\langle \lambda
_{RP}\right\rangle_{\lambda |R,\kappa}/\kappa\right\rangle
_{\kappa|R}$, $\left\langle \left\langle \lambda
_{PS}^2\right\rangle_{\lambda |P,\kappa}/\kappa \right\rangle
_{\kappa|P} = \left\langle \left\langle \lambda _{SR}^2
\right\rangle_{\lambda |S,\kappa}/\kappa \right\rangle_{\kappa|S}
= \left\langle \left\langle \lambda _{RP}^2 \right\rangle_{\lambda
|R,\kappa}/\kappa \right\rangle_{\kappa|R}$, and $\left\langle
\left\langle \lambda_{PS} \lambda_{RP}\right\rangle_{\lambda
|P,\kappa}/\kappa \right\rangle_{\kappa|P} = \left\langle
\left\langle \lambda _{RP}\lambda _{SR} \right\rangle_{\lambda
|R,\kappa}/\kappa \right\rangle_{\kappa|R}$.
%${\left\langle
%{\frac{{{{\left\langle {{\lambda _{PS}}} \right\rangle} _{\lambda
%|P,\kappa}}}}
% {\kappa }} \right\rangle _{\kappa|P}} = {\left\langle {\frac{{{{\left\langle {{\lambda
%_{RP}}} \right\rangle }_{\lambda |R,\kappa}}}}{\kappa }}
%\right\rangle _{\kappa|R}}$,
% ${\left\langle {\frac{{{{\left\langle {{\lambda _{PS}}^2} \right\rangle }_{\lambda |P,\kappa}}}}{\kappa }} \right\rangle _{\kappa|P}} =
%  {\left\langle {\frac{{{{\left\langle {{\lambda _{SR}}^2} \right\rangle }_{\lambda |S,\kappa}}}}{\kappa }} \right\rangle _{\kappa|S}}
%= {\left\langle {\frac{{{{\left\langle {{\lambda _{RP}}^2} \right
%\rangle }_{\lambda |R,\kappa}}}}{\kappa }} \right\rangle _{\kappa|R}}$, and
% ${\left\langle {\frac{{{{\left\langle {{\lambda _{PS}}{\lambda _{RP}}} \right\rangle }_{\lambda |P,\kappa}}}}{\kappa}} \right\rangle _{\kappa|P}} = {\left\langle {\frac{{{{\left\langle {{\lambda _{RP}}{\lambda _{SR}}} \right\rangle }_{\lambda |R,\kappa}}}}{\kappa}} \right\rangle _{\kappa|R}}$.
Secondly, the expected values can be expressed in terms of the
macroscopic quantities (link densities and fractions) that we want
to solve. The expected values ${{\left\langle {{{\left\langle
{{\lambda _{XY}}} \right\rangle }_{\lambda |X,\kappa}}}
\right\rangle }_{\kappa |X}}$ and $\langle \kappa
\rangle_{\kappa|X}$ are readily given by
\begin{equation}
\label{eq06}
\begin{array}{l}
{{\left\langle {{{\left\langle {{\lambda _{XY}}} \right\rangle
}_{\lambda |X,\kappa}}} \right\rangle}_{\kappa |X}}  =
\frac{\mu}{2f_{X}} \, l_{XY} \\
\langle \kappa \rangle_{\kappa|X}  = \frac{\mu}{2 f_{X}}
\left({{l_{XY}} + {l_{XZ}} + 2{l_{XX}}} \right) \,.
 \end{array}
\end{equation}
The first equation follows from $l_{XY} = \frac{2}{\mu N} n_{X}
\langle\lambda_{XY}\rangle_{X}$, where $n_{X}$ is the number of
agents using strategy-X. It says that the total number of
XY-links is given by the product of $n_{X}$ and the average
number of XY-links per agent using strategy-X.  The second
equation relates the mean degree $\langle \kappa
\rangle_{\kappa|X}$ among agents using strategy-X to the link
densities.

For agents using strategy-R of a certain degree $\kappa$, the
first moment ${\left\langle {\lambda _{RP}} \right\rangle
}_{\lambda |R,\kappa}$ and the second moment ${\left\langle
{\lambda _{RP}^2} \right\rangle_{\lambda |R,\kappa}}$ are related
to the expected value and the variance of $\lambda _{RP}$
respectively, and the mixed moment ${\left\langle {{\lambda _{RP}}
\cdot {\lambda_{SR}}} \right\rangle_{\lambda |R,\kappa}}$ is
related to the covariance of $\lambda _{RP}$ and $\lambda_{SR}$
via~\cite{stat1}
\begin{equation}
\label{moments}
\begin{array}{l}
\left\langle {\lambda_{RP}} \right\rangle_{\lambda |R,\kappa} =
\text{E}(\lambda_{RP}) \\
{\left\langle {\lambda_{RP}^2} \right\rangle_{\lambda |R,\kappa} =
\text{var}(\lambda_{RP}) + \left\langle \lambda_{RP}
\right\rangle_{\lambda |R,\kappa}^2} \\
{\left\langle {{\lambda _{RP}} \cdot {\lambda_{SR}}} \right\rangle
_{\lambda |R,\kappa} = \text{cov}( {\lambda _{RP}} , {\lambda_{SR}} ) + \left\langle
\lambda _{RP} \right\rangle_{\lambda |R,\kappa} \left\langle \lambda_{SR}
\right\rangle _{\lambda |R,\kappa}} \;.
\end{array}
\end{equation}
We invoke a trinomial closure scheme to handle
$\text{E}(\lambda_{RP})$, $\text{var}(\lambda_{RP})$ and
$\text{cov}({\lambda _{RP}},{\lambda_{SR}})$ and close the
equations.  It is an extension of the binomial closure scheme in
two-strategy models~\cite{vm,dasg2,ji,inverse}. The essence is to
treat averages $\langle \cdots \rangle_{\lambda |X,\kappa}$ that
involve the sums $ \sum_{\lambda_{XY},\lambda_{XZ}}
Q_{X,\kappa}(\lambda_{XY},\lambda_{XZ})(\cdots)$ approximately.
Physically, $Q_{X,\kappa}(\lambda_{XY},\lambda_{XZ})$ is the
probability of having exactly $\lambda_{XY}$ XY-links,
$\lambda_{XZ}$ XZ-links and $(\kappa - \lambda_{XY} -
\lambda_{XZ})$ XX-links, giving an agent with $\kappa$ neighbors
using the strategy-X.  This echoes the question on the
distribution of three possible outcomes $i=1,2,3$, each occurring
with the probability $p_{i}$, in $n$ independent trials.  The
resulting trinomial distribution gives the expected numbers
$np_{i}$ for the three outcomes, with the variances given by
$np_{i}(1-p_{i})$ and the covariances between different outcomes
$i$ and $j$ given by $-np_{i}p_{j}$~\cite{stat1,stat2}.  Here, the
degree $\kappa$ plays the role of $n$.  The probabilities $p_1$,
$p_2$ and $p_3$ are the conditional probabilities of encountering
a neighbor using strategies-R, P and S respectively, given
the strategy-X of an agent of degree $\kappa$.  Re-defining the
symbols of the probabilities as ${\rho _{R|X,\kappa }}$, ${\rho
_{P|X,\kappa }}$ and ${\rho _{S|X,\kappa }}$ respectively and
invoking the trinomial closure scheme, we have
\begin{equation}
\label{moments2}
\begin{array}{l}
\left\langle {\lambda_{RP}} \right\rangle_{\lambda |R,\kappa} =
\kappa \cdot {\rho _{P|R,\kappa }} \\
\left\langle {\lambda_{RP}^2} \right\rangle_{\lambda |R,\kappa} =
 \kappa \cdot {\rho _{P|R,\kappa }}\left( 1 - {\rho _{P|R,\kappa
 }}\right)
 + \left(\kappa \cdot \rho _{P|R,\kappa } \right)^{2} \\
\left\langle {{\lambda _{RP}} \cdot {\lambda_{SR}}} \right\rangle
_{\lambda |R,\kappa} = \kappa \cdot \rho _{P|R,\kappa} \cdot \rho
_{S|R,\kappa}\left( - 1 + \kappa \right) \;.
\end{array}
\end{equation}
To express all quantities in terms of the link densities, we make
the further approximation
\begin{equation}
\label{rhoY}
\rho _{Y|X,\kappa } = \frac{l_{XY}}{2
l_{XX}+l_{XY}+l_{XZ}}
\end{equation}
that the probability $\rho _{Y|X,\kappa}$ is given by the fraction
of out-going XY-links pointing to the neighbors using strategy-Y from all agents using strategy-X.  Note that this assumption
does not distinguish between different degrees $\kappa$ as $\rho
_{Y|X,\kappa}$ is independent of $\kappa$.

Finally, using Eqs.~(\ref{eq05}),~(\ref{moments2}), and
(\ref{rhoY}) allows us to express all the quantities in
Eq.~(\ref{eq02}) in terms of a link density and thus close the
equation.  The expected value in Eq.~(\ref{eq02}) vanish at long
time.  Setting the resulting equation to zero gives the link
density $l_{RP}$ as a function of the rewiring probability $p$.
The non-trivial solution of $l_{RP}(p)$ is found to be
\begin{equation}
\label{eq10} l_{RP}(p) = \frac{2}{9} \left( 1 - \frac{p}{3(\mu -
1)(1 - p)} \right) = \frac{2}{9 p_\text{cri}} \frac{ p_\text{cri}
- p}{1-p}
\end{equation}
for $p < p_\text{cri}$ with
\begin{equation}
\label{eq11} p_\text{cri} = \frac{3(\mu -1)}{3\mu - 2} \;,
\end{equation}
and $l_{RP}(p) = 0$ for $p > p_\text{cri}$.  Results for
$l_{RR}(p)$ follow form $l_{RR}(p) + l_{RP}(p) = 1/3$.

The analytic results in Eqs.~(\ref{eq10}) and (\ref{eq11}) are
shown in Fig.~\ref{fig02}(b) (lines) for comparison for the case
of $\mu=2$.  The results and the simulation results are in good
agreement.  The theory captures the two phases and the behavior of
the phase transition. There are slight discrepancies near the
phase transition. The theory predicts that $p_\text{cri} = 3/4$
for $\mu=2$, which is slightly lower than $p_\text{cri} \approx
0.78$ obtained by numerical simulations. The theory predicts a
shift in $p_\text{cri}$ to a higher value with increasing $\mu$,
which is a feature also observed in numerical simulations.
%{\bf Up to here}

%=======================================================
\section{Active agents win more than passive agents via co-evolving mechanism}
\label{sec4}

The discrepancies between analytic and simulation results, despite
small, reveals important information on the validity of the
assumptions in the mean-field approach and the effects of the
co-evolving mechanism, as we now show.  In every turn, the active
agent may win, lose, or draw.  Recording the probabilities of
winning, losing, and drawing of the active agents over many
rounds, the averages $\overline{f}_\text{win}$,
$\overline{f}_\text{draw}$ and $\overline{f}_\text{lose}$ of these
probabilities for the active agents can be obtained.  Due to the
cyclic symmetry of the strategies, we could focus on any strategy
for an active agent, say R, and express the three probabilities
as follows:
\begin{equation}
\label{eq12}
\begin{array}{l}
\overline{f}_\text{win} = {\left\langle \left\langle
{\frac{{{\lambda_{SR}}}}{\kappa }}
\right\rangle_{\lambda|R,\kappa} \right\rangle_{\kappa|R}} =
{\left\langle \rho _{S|R,\kappa } \right\rangle_{\kappa|R}} =
\left\langle f_{\text{win},\kappa}\right\rangle_{\kappa|R} \\
\overline{f}_\text{draw} = {\left\langle \left\langle
{\frac{{{\lambda_{RR}}}}{\kappa }}
\right\rangle_{\lambda|R,\kappa} \right\rangle_{\kappa|R}} =
{\left\langle \rho _{R|R,\kappa } \right\rangle_{\kappa|R}} =
\left\langle f_{\text{draw},\kappa}\right\rangle_{\kappa|R}\\
\overline{f}_\text{lose} = {\left\langle \left\langle
{\frac{{{\lambda_{RP}}}}{\kappa }}
\right\rangle_{\lambda|R,\kappa} \right\rangle_{\kappa|R}} =
{\left\langle \rho _{P|R,\kappa } \right\rangle_{\kappa|R}} =
\left\langle f_{\text{lose},\kappa}\right\rangle_{\kappa|R} \;.
\end{array}
\end{equation}
The quantities $\rho_{S|R,\kappa}$, $\rho_{R|R,\kappa}$ and
$\rho_{P|R,\kappa}$ were introduced in Eq.~(\ref{moments2}). They
are conditional probabilities of encountering a neighbor using
strategies-R, P and S respectively, given that the strategy
of the active agent is R and the degree is $\kappa$. In the
present context, they are also the probabilities of winning
($f_{\text{win},\kappa}$), drawing ($f_{\text{draw},\kappa}$) and
losing ($f_{\text{lose},\kappa}$) of an active agent who has a
degree $\kappa$.

\begin{figure}
\includegraphics[width=0.5\columnwidth]{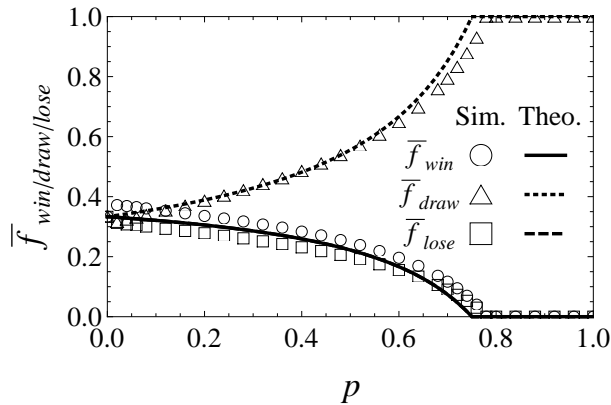}
\caption{Simulation (symbols) and mean-field results (lines) of
$\overline{f}_\text{win}$, $\overline{f}_\text{draw}$ and
$\overline{f}_\text{lose}$ in the steady state as a function of
$p$. The simulation data are obtained by averaging over 300
realizations in networks of $N=10000$.  The mean degree is
$\mu=2$.} \label{fig03}
\end{figure}

Fig.~\ref{fig03} shows the numerical results of these
probabilities as a function of $p$ for the case of mean degree
$\mu=2$. These results are illuminating. At $p=0$,
$\overline{f}_\text{win} = \overline{f}_\text{draw} =
\overline{f}_\text{lose} = 1/3$. A slight deviation from $p=0$
immediately makes $\overline{f}_\text{win }$, $
\overline{f}_\text{draw}$ and $\overline{f}_\text{lose}$ different
from 1/3 with a jump. For $p>0$, these quantities also illustrate
the existence of two phases. In the active phase,
$\overline{f}_\text{win}$ and $\overline{f}_\text{lose}$ drops
monotonically with $p$ and vanish for $p > p_\text{cri}$, while
$\overline{f}_\text{draw}$ increases monotonically with $p$ and
becomes unity for $p > p_\text{cri}$. The most important feature
is $\overline{f}_\text{win}
> \overline{f}_\text{lose}$ for active agents in the active phase,
i.e., active agents are more likely to win on average. In
contrast, passive agents are more likely to loss on average. Thus,
examining the numerical results of $\overline{f}_\text{win }$, $
\overline{f}_\text{draw}$ and $\overline{f}_\text{lose}$ indicates
a deficiency in the theory.  The theory assumes
$\overline{f}_\text{win} = \overline{f}_\text{lose}$ and
approximates them by $\left\langle \rho _{S|R,\kappa}
\right\rangle_{\kappa|R} = \left\langle \rho _{P|R,\kappa}
\right\rangle_{\kappa|R} = 3l_{RP}/2$ (see Eq.~(\ref{rhoY})). The
analytic results of $\overline{f}_\text{win}$,
$\overline{f}_\text{lose}$ and $\overline{f}_\text{draw}$ are also
shown in Fig.~\ref{fig03} (lines) for comparison.  For a large
part of $p$ below $p_\text{cri}$, the analytic results lie between
the actual $\overline{f}_\text{win}$ and
$\overline{f}_\text{lose}$. However, for $p \lesssim
p_\text{cri}$, the analytic results go below both
$\overline{f}_\text{win}$ and $\overline{f}_\text{lose}$. The
analytic results are in exact agreement with the simulation
results right at $p=0$, but do not predict the jump in
$\overline{f}_\text{win}$, $\overline{f}_\text{draw}$ and
$\overline{f}_\text{lose}$ for any deviation from $p=0$.

\begin{figure}
\subfigure{%
\includegraphics[width=0.45\columnwidth]{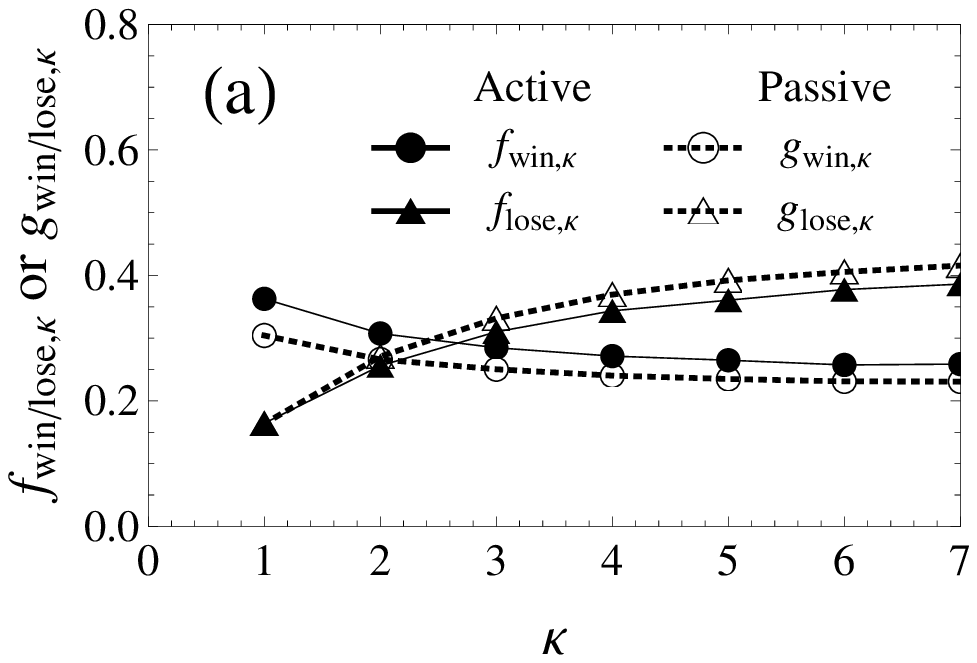}
}%
\subfigure{%
\includegraphics[width=0.45\columnwidth]{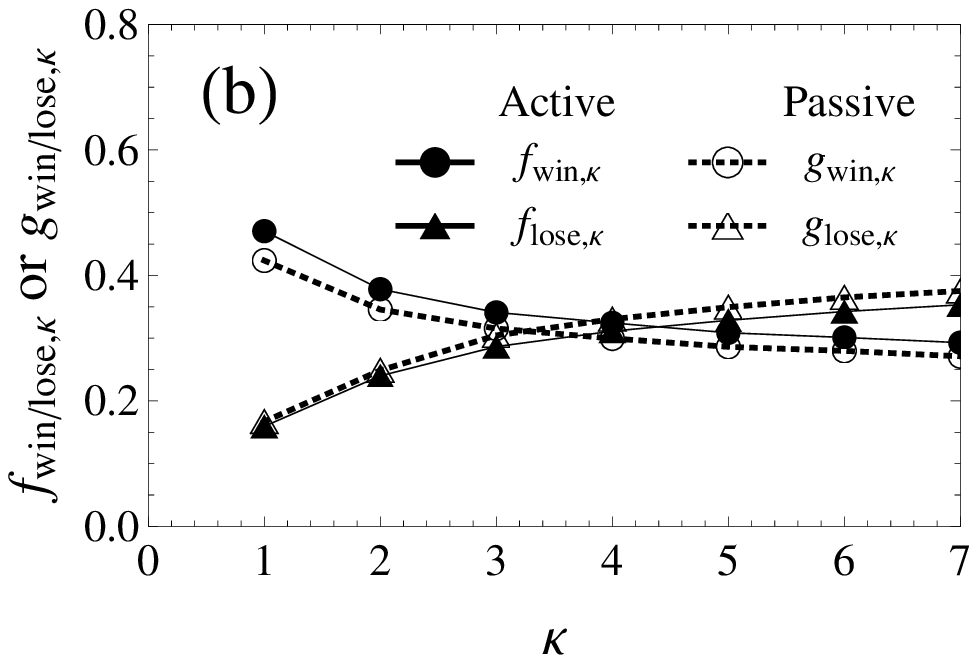}
}%
\caption{Simulation results of $f_{\text{win},\kappa}$,
$f_{\text{lose},\kappa}$, $g_{\text{win},\kappa}$ and
$g_{\text{lose},\kappa}$ in the steady state for agents of
different degrees $\kappa$ at $p=0.3$ for systems with mean degree
(a) $\mu=2$ and (b) $\mu=4$. The data are obtained by averaging
results of 300 independent runs in a network of $N=10000$.  The
lines joining the data points serve as a guide to the eyes.}
\label{fig04}
\end{figure}

We now discuss the validity of the mean field theory in light of
these features. In the theory, the quantity $\rho_{Y|X,\kappa}$,
which is the fraction of links to neighbors using strategy-Y for
agents using strategy-X and having $\kappa$ neighbors, is
approximated by Eq.~(\ref{rhoY}) and assumed to be independent of
$\kappa$.  Thus, $f_{\text{win},\kappa}$, $f_{\text{lose},\kappa}$
and $f_{\text{lose},\kappa}$ are also assumed to be independent of
$\kappa$.  At $p=0$, there is no rewiring. The network is static
and every agent has the same number $\mu$ of neighbors. The fact
that the theory gives the correct value at $p=0$ but not for $p
\neq 0$ implies that the spread in the values of $\kappa$ becomes
important when rewiring is present. Indeed, agents acquire
different values of $\kappa$ due to the rewiring mechanism.
Fig.~\ref{fig04} shows the simulation results of
$f_{\text{win},\kappa}$ and $f_{\text{lose},\kappa}$ as a function
of $\kappa$ at a fixed $p=0.3$ for two different systems of
$\mu=2$ and $\mu=4$.  We also recorded the winning and losing
probabilities $g_{\text{win},\kappa}$ and $g_{\text{lose},\kappa}$
of passive agents of degree $\kappa$ and showed the results. It is
important to note that $f_{\text{win},\kappa}$ and
$f_{\text{lose},\kappa}$ do depend on $\kappa$, and so do
$g_{\text{win},\kappa}$ and $g_{\text{lose},\kappa}$.  This
dependence on $\kappa$, which enters for any $p \neq 0$, causes
the mean field theory to miss the jump in the probabilities as $p$
starts to take on finite values (see Fig.~\ref{fig03}).

Closer inspection of the results in Fig.~\ref{fig04} reveal that
$f_{\text{win},\kappa} > f_{\text{lose},\kappa}$ and
$g_{\text{win},\kappa} > g_{\text{lose},\kappa}$ for $\kappa <
\mu$; but $f_{\text{win},\kappa} < f_{\text{lose},\kappa}$ and
$g_{\text{win},\kappa} < g_{\text{lose},\kappa}$ for $\kappa >
\mu$.  Although the results in Fig.~\ref{fig04} were obtained for
$p=0.3$, we examined the range of $0<p<p_{\text{cri}}$ and found
the same features. Thus, an active or passive agent with a degree
smaller (larger) than the mean degree $\mu$ is more likely to win
than to lose (to lose than to win) in a RPS game, while the
probabilities of winning and losing of an active or passive agent
who has a degree $\kappa \approx \mu$ are nearly identical. More
importantly, $f_{\text{win},\kappa}>g_{\text{win},\kappa}$ and
$f_{\text{lose},\kappa}<g_{\text{lose},\kappa}$ for all $\kappa$,
implying that an active agent is more likely to win than a passive
agent of the same $\kappa$.

The $\kappa$-dependence of the winning and losing probabilities
can be understood qualitatively as follows.  An agent takes
actions to make his neighborhood better, i.e., to enhance his
chance of winning.  Switching strategy helps an active agent to
win over the same opponent if they meet again (provided that their
strategies are not further altered before they meet again).
Rewiring dissatisfying link lowers the losing probability of an
active agent when he becomes involved in a RPS game later.
Generally, the neighborhood of an active (a passive) agent gets
better (gets worse) after an adaptive action takes place. The
probability of an agent to be chosen as an active agent in a time
step is $1/N$. However, the probability of an agent being a
passive agent depends on his degree $\kappa$. Ignoring spatial
correction in the network for simplicity, the probability of being
a passive agent is $\kappa/\mu N$, as given by the ratio of his
out-links to the total number of out-links in the network.  Here,
the ratio $\kappa/\mu$ emerges.  For agents with $\kappa < \mu$,
they are more likely to be active agents and thus a better chance
to shape his neighborhood to his advantage.  The more favorable
neighborhood gives them a larger winning probability than losing
in the next RPS game, no matter which role they play. Therefore,
the co-evolving mechanism leads to $f_{\text{win},\kappa} >
f_{\text{lose},\kappa}$ and $g_{\text{win},\kappa} >
g_{\text{lose},\kappa}$ for $\kappa < \mu$, as shown in
Fig.~\ref{fig04}.  Following a similar argument, agents with
$\kappa > \mu$ are more likely to be passive agents. On one hand,
they do not have much chance to make his neighborhood better.  On
the other hand, their neighbors' adaptive actions make the
neighborhood worse.  These agents will have a higher losing
probability than winning.  Therefore, the co-evolving mechanism
leads to $f_{\text{win},\kappa} < f_{\text{lose},\kappa}$ and
$g_{\text{win},\kappa} < g_{\text{lose},\kappa}$ for $\kappa >
\mu$, also observed in Fig.~\ref{fig04}.  The physical picture is
that agents with many neighbors (high $\kappa$) are those often
defeated by their neighbors and so the neighbors want to keep the
relationship, while agents with only a few neighbors can protect
themselves from losing and strive for higher chance of winning in
the next RPS game.

The analysis on how $f_{\text{win},\kappa}$ and
$f_{\text{lose},\kappa}$ depend on $\kappa$ brings out the
inadequacy of the mean field theory in capturing the spatial
correlation between neighboring agents' strategies after the
system evolves to a steady state.  The results in Fig.~\ref{fig04}
imply $\rho_{S|R,\kappa}>\rho_{P|R,\kappa}$ for $\kappa<\mu$ while
$\rho_{S|R,\kappa}<\rho_{P|R,\kappa}$ for $\kappa>\mu$.  Such
correlations are not captured by the approximation of
$\rho_{Y|X,\kappa} $ by Eq.~(\ref{rhoY}).  This inadequacy also
leads to the discrepancies in evaluating $l_{RP}$ and $l_{RR}$ for
$0<p<p_{\text{cri}}$, in addition to $\overline{f}_\text{win}$,
$\overline{f}_\text{draw}$ and $\overline{f}_\text{lose}$.
Finally, the analysis in Fig.~\ref{fig04} provides an
understanding of why the averaged probabilities
$\overline{f}_\text{win}> \overline{f}_\text{lose}$, as shown in
Fig.~\ref{fig03}. It is a combined effect of (i) a randomly
selected neighbor (passive agent) is expected to have a higher
degree than a randomly selected agent (active agent)
\cite{friend}, and (ii) $f_{\text{win},\kappa}>
g_{\text{win},\kappa}$
($f_{\text{lose},\kappa}<g_{\text{lose},\kappa}$) and they
decrease (increase) monotonically as $\kappa$ increases (see
Fig.~\ref{fig04}).

\section{Conclusion}
\label{sec5}
%=======================================================
To summarize, we have proposed and studied an adaptive
Rock-Paper-Scissors model (ARPS) in detail, with a focus on issues
related to formulating a mean field theory for co-evolving network
problems with multiple strategies.  In ARPS, three cyclically
dominating strategies are involved in a co-evolving network.  An
agent with a dissatisfied neighbor takes action to improve his
competing neighborhood by rewiring the dissatisfying link with a
probability $p$ or switching to a strategy that could defeat the
neighbor with a probability $(1-p)$.  The network shows two
different phases: an active phase for $p<p_{\text{cri}}$ and a
frozen phase for $p>p_{\text{cri}}$. The active phase is
characterized by one connected network with agents using different
strategies continually interacting and taking adaptive actions.
The frozen phase is characterized by three separate clusters of
agents using R, P, and S, respectively and terminated adaptive
actions.  We have discussed in detail the formulation of a
mean-field theory that starts with tracing the changes in a link
density due to all possible adaptive actions as the network
evolves.  A trinomial closure scheme, which approximates the
distribution of different types of lines that an agent carries
given his strategy and degree, has been invoked to close the
equation.  Ignoring the dependence on the degree, the theory gives
an analytic expression for the link density as a function of $p$.
The results agree with simulation results well and capture the
two-phase structure.

Closer examination of the small deviations between analytic and
simulation results turns out to be illuminating. We have studied
the averaged probabilities of winning ($\overline{f}_\text{win}$),
drawing ($\overline{f}_\text{draw}$) and losing
($\overline{f}_\text{lose}$) for active agents. It was found that
$\overline{f}_\text{win}$ is always higher than
$\overline{f}_\text{lose}$ in the active phase - a feature that
the mean field theory does not capture.  The origin has been
traced to the spread in the degrees among agents due to rewiring
{\em and} the dependence of the winning and losing probabilities
on the degree of agents.  We have found that agents with a degree
smaller (larger) than the mean degree $\mu$ have a larger
(smaller) probability of winning than losing.  Physically, active
agents tend to have smaller degrees than passive agents because
links are retained or increased only for agents who are being
taken advantage of.  The results are useful in that the inclusion
of correlations between the nearest neighbors' strategies and
degrees should give a more accurate theory.

We close with a discussion on a few possible extensions.  In the
present work, we simplified the discussion by using the symmetry
that comes from the cyclically dominating strategies as well as
the random initial strategy assignments.  It will be interesting
to study the sensitivity of the steady state to different initial
strategy distributions.  The theory presented here can also be
modified to study the problem.  Here, we discussed the analytic
approach not only for applying the results to ARPS, but also in a
general way that could be readily modified to other co-evolving
network models involving multiple strategies.  These models need
not be cyclically dominating and the number of strategies could be
more than three.  The detailed study on the reasons of the small
deviation between analytic and simulation results provides useful
information on how better theories can be formulated. The analytic
results also provides a guide for further studies on the scaling
behavior near the transition between the two phases.

%\section{acknowledgements}
%\label{acknowledgement}

\end{document}